\documentstyle[11pt]{article}
  \newcommand {\nc}{\newcommand}
  \nc{\eq}{\begin{equation}}
  \nc{\en}{\end{equation}}
  
  \def\intg{{\cal Z}}

  \def\Alg{{\cal A}}
  \def\Hs{{\cal H}}
  \def\Drc{{\cal D}}
  \def\Lh{{\cal L}}
  \def\etal{{\it et al} }
 \title{Connes' Distance of One-Dimensional Lattices: \\
 General Cases
 }
 \author{
  Jian DAI$^\dag$, Xing-Chang SONG\\
  Theory Group, Department of Physics, \\Peking University, Beijing, P.R.China, 100871\\
  \dag Room 2082, Building 48, \\Peking University, Beijing, P. R. China,
  100871\\
  E-mail: jdai@mail.phy.pku.edu.cn, songxc@ibm320h.phy.pku.edu.cn
  }
 \date{March 19th, 2001}
 
\begin{document}
  \maketitle
  \begin{abstract}
  \noindent
   Connes' distance formula is applied to endow linear metric to three 1D lattices of different
   topology, with a generalization of lattice Dirac operator
   written down by Dimakis \etal to contain a non-unitary
   link-variable. Geometric interpretation of this link-variable is
   lattice spacing and parallel transport.
  \end{abstract}
  \section{Introduction}
   Lattice as a universal regulator for the non-perturbative definition of a quantum field
   theory works well for bosonic fields \cite{Wilson}. However,
   when fermionic fields are involved, lattice formalism
   encounters two well-known seemingly insurmountable problems:
   implementation of grassmann number in simulations and No-Go theorem
   for chiral fermion on lattices \cite{Creutz}. On the other hand,
   lattice provides one simplest model of noncommutative
   geometry(NCG) \cite{dimakis_L}; NCG in Connes' formulation
   has an intimate relation with fermion through a Hilbert space and a generalized Dirac
   operator \cite{NCG_F}. Therefore, to explore lattice field
   theory in NCG context is significant for to understand those old
   puzzles. As the first step, because NCG endows a metric, hence a geometry, onto a space through Dirac
   operator, to consider this (Dirac-operator)induced metric on lattices
   exhibits new relation between lattice fermions and lattice geometry. In fact, the first striking nontrivial result along this line is
   that this distance is non-Euclidean, providing Na\"{\i}ve or Wilson-Dirac operator is adopted
   \cite{Bimonte}\cite{Atzmon}. On the contrary, Dimakis and M\"{u}ller-Hoissen(DM) proposed a new free Dirac operator which induces correct linear
   distance on a 1D lattice \cite{Dimakis_D}. In this paper, we generalize
   DM's result in case that a {\it link-variable} field is presented on this 1D
   lattice. We will show that the amplitude of this field modify
   the induced distance in the sense that its inverse provides a localized
   lattice spacing and that the phase of this field can play the
   role of a $U(1)$-parallel transport, hence a gauge potential.\\

   This paper is organized as following. Connes' distance is introduced in
   Sect.\ref{sec2}, and is calculated for three types of 1D lattices in
   Sect.\ref{sec3} after generalized DM's lattice Dirac operator is defined.
   Geometric interpretation is given in Sect.\ref{sec4}.
  \section{Connes' Distance Formula}\label{sec2}
   A spectral geometry in Connes' sense, commutative or not, is defined to be a triple
   $(\Alg,\Hs,\Drc)$ in which $\Alg$ is a pre-$C^\ast$ algebra
   being represented faithfully on Hilbert space $\Hs$ and $\Drc$
   is a self-adjoint operator on $\Hs$ playing the role of Dirac
   operator in Classical spinor geometry \cite{IntrodNCG}. In this paper,
   $\Alg$ is taken to be the algebra of complex functions on a
   lattice, $\Hs$ is the Hilbert space of fermionic fields which
   are not considered as grassmann-valued sections, and $\Drc$ is lattice Dirac
   operator to be specified. Connes' distance is introduced by
   the formula
   \eq\label{dist}
    d_\Drc(p,q)=sup_f\{|f(p)-f(q)|:f\in \Alg, ||[\Drc,f]||_\Hs\leq 1\}
   \en
   for all points $p,q$ of this lattice, where $||.||_\Hs$ is
   operator norm on $\Lh(\Hs)$. Note that we do not distinguish
   $f$ from its imagine represented on $\Hs$ due to the
   faithfulness. To obtain a manipulable algorithm for
   Eq.(\ref{dist}), we define a $f${\it -Hamiltonian},
   $H(f)=[\Drc,f]^\dag[\Drc,f]$.
   Then it is easy to verify that
   $||[\Drc,f]||_\Hs^2=sup_{\lambda_\Drc(f)}\{\lambda_\Drc(f):H(f)\psi=\lambda_\Drc(f)\psi\}$.
   Consequently, Eq.(\ref{dist}) can be expressed as
   \eq\label{dist1}
    d_\Drc(p,q)=sup_f\{|f(p)-f(q)|:f\in \Alg, \forall \lambda_\Drc(f)\leq
    1\}
   \en
  \section{Lattice Dirac Operator and Induced
  Distance}\label{sec3}
   We specify the term {\it one-dimensional lattice} by a discrete
   set $L$ together with a isomorphism $T$ acting on $L$.
   $\Alg(L)$ is denoted for algebra of complex functions on $L$
   and Hilbert space is chosen to be
   $\Hs=\Alg(L)\oplus\Alg(L)$ which is a free module over $\Alg(L)$ of rank
   2. $T$ induces an isomorphism of $\Alg(L)$ and an isometry of $\Hs$ to which we still write as $T$.
   DM's free lattice Dirac operator can be written as
   \[
    \Drc=T\sigma^++T^\dag\sigma^-
   \]
   where $\sigma^\pm$ are defined using Pauli matrices $\sigma^\pm=(\sigma_1\pm i\sigma_2)/2$. We
   generalize it to be
   \eq\label{Drc}
    \Drc(\omega)=\omega T\sigma^++T^\dag\bar{\omega}\sigma^-
   \en
   where $\omega\in\Alg(L)$. Below we consider three types of $(L,T)$ corresponding to three topologies in continuum limit.
   \subsection{Finite Open Lattice $\hat{\intg}_N$}\label{subsec}
    In this case, $L$ is coordinatized by $1,2,...,N$ and
    $(Tf)(i)=f(i+1),i=1,2,...,N-1;(Tf)(N)=0$ for all $f\in\Alg(L)$, which we refer as $\hat{\intg}_N$.
    Notice Eq.(\ref{Drc}), $[\Drc(\omega),f]=\omega\partial^+ f
    T\sigma^++T^\dag\bar{\omega}(-\partial^+ f)\sigma^-$, where
    $(\partial^+ f)(i)=(Tf)(i)-f(i),i=1,2,...,N-1;(\partial^+ f)(N)=0$.
    One can check that $f$-Hamiltonian
    $H(f)=|\omega|^2|\partial^+ f|^2\sigma^+\sigma^-+|T^\dag\omega|^2|\partial^- f|^2\sigma^-\sigma^+$ where
    $(\partial^-f)(i+1)=f(i)-(Tf)(i),i=1,2,...,N-1;(\partial^-f)(1)=0$.
    Therefore,
    $||[\Drc(\omega),f]||_\Hs^2=||(|\omega\partial^+f|)^2||_\infty$
    in which $||.||_\infty$ is {\it sup}-norm of $\Alg(\hat{\intg}_N)$.
    According Eq.(\ref{dist}),
    \[
     d_{\Drc(\omega)}(i,j)=sup_f\{|f(i)-f(j)|:f\in\Alg(\hat{\intg}_N),|\omega\partial^+f|(k)\leq 1,k=1,2,...,N\}
    \]
    for all $i,j\in \hat{\intg}_N$. If we assume $\omega$ is
    non-singular, i.e. $|\omega(k)|\neq 0$ for all $k$, then
    $d_{\Drc(\omega)}(i,j)$ possesses an upper bound
    \eq\label{bd}
     d_{\Drc(\omega)}(i,j)\leq
     \sum_{k=0}^{j-i-1}|\omega(i+k)|^{-1}
    \en
    in which $j$ is supposed to be larger than $i$. Define
    $f_\omega(i+1)=f_\omega(i)+|\omega(i)|^{-1},f_\omega(1)=0$, then
    $||[\Drc(\omega),f_\omega]||\leq 1$ and
    $f_\omega$ saturates the upper bound in Eq.(\ref{bd}).
    Subsequently, (\ref{bd}) becomes an equality, especially it holds that
    $d_{\Drc(\omega)}(i,i+1)=|\omega(i)|^{-1}$, to which a clear
    interpretation is that the inverse of amplitude of $\omega(i)$
    is the lattice spacing between $i$ and $i+1$. Note that it is
    obvious that the value of $\omega$ at $N$ makes no sense in this case.
   \subsection{Finite Close Lattice $\intg_N$}
    Here $L$ is labeled by $0,1,2,...,N-1$ and
    $(Tf)(i)=f(i+1),i=0,1,...,N-2;(Tf)(N-1)=f(0)$ for all $f\in\Alg(L)$, so addition of the argument of $f$ makes $L$ a finite group $\intg_N$.
    If we define $\partial^+f=Tf-f,\partial^-f=T^\dag f-f$, then
    the deduction is exactly the same as that in
    Subsect.\ref{subsec} and
    \[
     d_{\Drc(\omega)}(i,j)=sup_f\{|f(i)-f(j)|:f\in\Alg(\intg_N),|\omega\partial^+f|(k)\leq
     1,k\in\intg_N \}, \forall i,j\in\intg_N
    \]
    With the non-singular assumption on $\omega$ and cyclic addition on $\intg_N$,
    \eq\label{bd1}
     d_{\Drc(\omega)}(i,j)\leq min\{l(i,j),l(j,i)\}
    \en
    where
    $l(i,j)=|\omega(i)|^{-1}+|\omega(i+1)|^{-1}+...+|\omega(j-1)|^{-1}$.
    Now we design a function to saturate this upper bound. Without
    losing generality, let $l(i,j)\leq l(j,i)$ and define
    $f_\omega(i)=0,f_\omega(i+1)=|\omega(i)|^{-1},f_\omega(i+2)=f_\omega(i+1)+|\omega(i+1)|^{-1},
    ...,f_\omega(j)=f_\omega(j-1)+|\omega(j-1)|^{-1},f_\omega(j+1)=f_\omega(j)-|\omega(j)|^{-1}l(i,j)l(j,i)^{-1},
    f_\omega(j+2)=f_\omega(j+1)-|\omega(j+1)|^{-1}l(i,j)l(j,i)^{-1},...,f_\omega(i-1)=f_\omega(i-2)-|\omega(i-2)|^{-1}l(i,j)l(j,i)^{-1}$.
    It is easy to check that $||[\Drc(\omega),f_\omega]||\leq 1$ and that $f_\omega$ saturates the upper bound in
    Eq.(\ref{bd1}). If $\omega$ satisfies triangle-inequalities $|\omega(i)|^{-1}\leq
    \sum_{k=1}^{N-1}|\omega(i+k)|^{-1}, \forall i\in\intg_N$, then $|\omega(i)|^{-1}$
    is able to be interpreted as lattice spacing between $i$ and
    $i+1$.
   \subsection{Infinite Lattice $\intg$}
    $L$ is parametrized by integer $\intg$ in this case and
    $(Tf)(i)=f(i+1), \forall i\in\intg,f\in\Alg(L)$. However to guarantee convergency, we
    must consider $\Hs=l^2(\Alg(L)\oplus\Alg(L))$ and $\Alg(\intg)=\{f\in\Alg(L):||[\Drc(\omega),f]||_\Hs<\infty\}$ here.
    Still define $\partial^+f=Tf-f,\partial^-f=T^\dag f-f$, then
    deduction is the same as that in Subsect.\ref{subsec}, and it
    follows that
    \[
     d_{\Drc(\omega)}(i,j)=sup_f\{|f(i)-f(j)|:f\in\Alg(\intg),|\omega\partial^+f|(k)\leq
     1,k\in\intg \}, \forall i,j\in\intg
    \]
    With non-singular $\omega$ and that $i<j$,
    \eq\label{bd2}
     d_{\Drc(\omega)}(i,j)\leq
     \sum_{k=0}^{j-i-1}|\omega(i+k)|^{-1}
    \en
    Let $f_\omega(0)=0,
    f_\omega(k)=f_\omega(k-1)+|\omega(k-1)|^{-1},f_\omega(-k)=f_\omega(-k+1)-|\omega(-k)|^{-1},
    k=1,2,...$, then $||[\Drc(\omega),f_\omega]||\leq 1$ and $f_\omega$ saturates the upper bound in
    Eq.(\ref{bd2}). Since
    $d_{\Drc(\omega)}(i,i+1)=|\omega(i)|^{-1}$, $|\omega(i)|^{-1}$
    is the lattice spacing between $i$ and $i+1$.\\

    Notice that non-singular $\omega$ can be polarized as
    $a_+^{-1}e^{ia_+A}$ with two real functions $a_+,A$, we conclude that $d_{\Drc(\omega)}$ is determined entirely by lattice spacing
    function $a_+$ and that $d_{\Drc(\omega)}$ is still linear
    distance in the sense of additivity.
  \section{Discussions}\label{sec4}
   We claim that $e^{ia_+A}$ in the above decomposition plays the role of unitary link-variable
   in lattice gauge theory, or equivalently parallel transport in mathematical literature.
   In fact, a local $U(1)$-gauge transformation on $\Hs$ is defined to be
   $\psi\rightarrow u\psi, \forall
   \psi\in\Hs$ where $u$ is a unitary in $\Alg(L)$ and
   a $U(1)$-parallel transport $U$ on $L$ is a link-variable satisfying $U\rightarrow uU\bar{u}$, $<U\psi,U\psi>=<T\psi,T\psi>$
   in which $<,>$ is hermitian-structure on $\Hs$. If $e^{ia_+A}\rightarrow ue^{ia_+A} (T\bar{u}), a_+\rightarrow a_+$, then
   $e^{ia_+A}T$ is a parallel transport and $(\psi,\Drc_\omega\psi)$ is gauge-invariant where $(,)$ is inner
   product of $\Hs$. Therefore geometric interpretation of
   $\omega$ is clear: $\omega$ is a link-variable not necessarily
   unitary, whose amplitude provides a {\it vierbein} and
   phase is the usual integrated $U(1)$-connection.\\

   Non-unitary link-variable has been noticed in the work of Majid and Raineri \cite{Majid} discussing field theory on
   permutation group $S_3$ and ours \cite{DS_W}. Nevertheless, its geometric picture is the clearest on
   1D lattices.\\

   {\bf Acknowledgements}\\
    This work was supported by Climb-Up (Pan Deng) Project of
    Department of Science and Technology in China, Chinese
    National Science Foundation and Doctoral Programme Foundation
    of Institution of Higher Education in China. We are grateful to Prof. S. Majid
    for introducing his work to us.
  
 \end{document}